\documentstyle{article}
\newcommand{\be}{\begin{equation}}
\newcommand{\ee}{\end{equation}}
\def\bea{\begin{eqnarray}}
\def\eea{\end{eqnarray}}
\begin{document}
\begin{titlepage}
\begin{flushright}    UFIFT-HEP-98-22 \end{flushright}
%\begin{frontmatter}
\vskip .8cm
\centerline{\LARGE{\bf {M(ysterious) Patterns in $SO(9)$}}}
%\title{M(ysterious) Patterns in $SO(9)$}
\vskip 1.5cm
\centerline{\bf Tekparson Pengpan, and Pierre Ramond}   
\vskip .5cm
\centerline{\em  Institute for Fundamental Theory,}
\centerline{\em Department of Physics, University of Florida}
\centerline{\em Gainesville FL 32611, USA}
\vskip 1cm
%\author{Tekparson Pengpan, Pierre Ramond}
%\address{Institute for Fundamental Theory, 
%Department of Physics, University of Florida, Gainesville FL 32611, USA}
\centerline{\bf {Abstract}}
\vskip 2cm
\noindent 
%\begin{abstract}
The light-cone little group, $SO(9)$, classifies the massless degrees of freedom  of eleven-dimensional supergravity, with a triplet of  representations.  We observe that this triplet generalizes to  four-fold infinite families with the quantum numbers of 
massless higher spin states. Their mathematical structure stems from the three 
equivalent ways of embedding $SO(9)$ into the exceptional group $F_4$.  
\vskip 1in
\centerline{\it To be published in a special issue of Physics Reports}
\centerline{\it in memory of Richard Slansky}
%\end{abstract}
%\end{frontmatter}
\end{titlepage}
\section{N$=1$ Supergravity in Eleven Dimensions}
It has been recently pointed out that $11$-dimensional supergravity is 
the local limit of a much bigger theory, called M-theory~\cite{MT}, that  also contains in different limits all known string theories in ten dimensions. At present, it is still elusive, and only a partial formulation~\cite{MATRIX} exists in the literature. 

Since M-theory lives in eleven dimensions, its massive degrees of freedom must be expressible as multiplets of $SO(10)$, the Lorentz little group of eleven dimensions and its massless degrees of freedom must form in  representations of $SO(9)$. Among those are the fields of the local supergravity theory which reveal themselves in the local limit. While it is likely that some of the physical objects in M-theory are not local, one still expects that they would be expressible in terms of infinite towers of representations of these little groups. 

There is a  pervading lore against interacting theories that contain massless states of higher spin. It is based on several no-go theorems,  formulated in terms of local field theory~\cite{RANDY}. They state that relativistically invariant theories with a finite number of local massless fields of spin higher than two, and with a finite number of derivatives in their interactions, do not exist. It follows that any such theory with an infinite tower of fields, and arbitrarily high derivative couplings escapes the no-go theorems and could conceivably exist.  Even with supersymmetry, building such a theory seems like a hopeless task, and the many published attempts  have met with partial success. A four-dimensional formulation~\cite{Vasiliev} uses an infinite-dimensional superalgebra, with the interesting feature that it necessarily contains a cosmological constant. Hence it seems that the lore against massless high-spin interacting theories is mainly based on the difficulties associated with their construction rather than on their impossibility. Since M-theory is most likely non-local, it may evade the no-go theorems, and could contain an infinite number of fields. It is therefore interesting to examine the $SO(9)$ properties of eleven-dimensional supergravity, whose massless states are  local limit of M-theory. 

In the following, we would like to draw attention to a remarkable mathematical fact, which shows that the supergravity triplet of $SO(9)$ representations  is actually the tip of a mathematical iceberg. We will start by presenting group-theoretical evidence that the supergravity representations are the first of an infinite family of  massless states of higher spin. Then we will offer a mathematical resolution in terms of embeddings of $SO(9)$ into the exceptional group $F_4$, as well as some generalizations. Since there are no coincidences in the study of these highly constrained theories, it is tempting to muse that these extra higher-spin massless states represent the degrees of freedom of M-theory, even though we have not been able to obtain any  dynamical evidence for this conjecture. 

\section{ Group Phenomenology of $SO(9)$}
The classical Lie group $SO(9)$ plays an important  dual role in the study of  theories in ten and eleven dimensions, as the light-cone little group of  Lorentz-invariant theories in ten space and one time dimensions, and as  the little group of massive representations of theories in nine space and one time dimensions. 

The representations of SO(9) are best described in Dynkin's language, which Dick Slansky 
used to great effectiveness in particle physics~\cite{SLANSKY}. As a rank $4$ Lie algebra, it takes four positive integers to label its irreducible representations, in the form $[a_1~a_2~a_3~a_4]$. Its four basic representations are:
\begin{itemize}
\item Vector, $[1000]$, with $9$ components, $V_i$,
\item Adjoint,  $[0100]$, with $36$ components, $B_{[ij]}$,
\item Three-form,  $[0010]$, with $84$ components, $B_{[ijk]}$,
\item Spinor,  $[0001]$, with $16$ components, $\psi_\alpha$.
\end{itemize}
All representations with odd $a_4$ are spinorial. The irreps of $SO(9)$ are characterized by five generalized Dynkin indices
\be
I_p\equiv\sum_{\rm rep}w^p\ ,\ \ p=0,2,4,6,8\ ,\ee
where $w$ are the weights in the representation. Thus $I_0$ is the dimension of the irrep, and  $I_2$ is related to the quadratic Casimir invariant by $C_2=36I_2/I_0$.

$N=1$ supergravity in eleven dimension is a local field theory that contains three different massless fields, two bosonic that describe gravity and a three-form, and one Rarita-Schwinger spinor. Its physical degrees of freedom are classified in terms of the light-cone little group, $SO(9)$, 
\begin{itemize}
\item Graviton as a symmetric second-rank tensor, $[2000]$, $G_{(ij)}$,
\item Third-rank antisymmetric tensor, $[0010]$, $B_{[ijk]}$,
\item Rarita-Schwinger spinor-vector, $[1001]$, $\Psi_{\alpha i}$.
\end{itemize} 
Their group-theoretical properties are summarized in the following table~\cite{PATERA}
\hskip 2cm
\begin{center}
\begin{tabular}{|c|c|c|c|}
\hline
$~{\rm irrep}~$& $[1001]$&$ [2000]$ & $[0010]$   \\
 \hline \hline         
$~I_0~ $&$  128$ & $ 44$ & $ 84$  \\
 \hline    
$~I_2~$& $256$& $88$ & $168$  \\
\hline
$~I_4~$& $640$& $232$ & $408$ \\                                                            
 \hline
$~I_6~$&$1792$& $712$ &$1080$\\ 
\hline
$~I_8~$&$5248$& $2440$ &$3000$\\ 
\hline
 \end{tabular}\end{center}
\vskip 0.3cm
We note that these indices, except for $I_8$, match between the fermion and the two bosons. 
As is well known, equality of the bosonic and fermionic dimensions is an indication of supersymmetry. On the light-cone, the supersymmetry algebra reduces to 
\be \{Q_\alpha,Q_\beta\}=\delta_{\alpha\beta}\ ,\ee
where the supersymmetric generators transform as the $\bf 16$ spinor of 
$SO(9)$. They split into creation and annihilation operators under the decomposition
\be SO(9)\supset SO(6)\times SO(3)\ ;\qquad
 {\bf 16}=({\bf\overline 4},{\bf 2})+({\bf 4},{\bf 2})\ ,\ee
and we obtain a Clifford algebra
\be \{\tilde Q,\tilde Q^\dagger\}=1\ ,\ee
where
$\tilde Q$ transforms as $({\bf\overline 4},{\bf 2})$, and $\tilde Q^\dagger$ 
as $({\bf 4},{\bf 2})$. The states of the Sugra multiplet are then 
obtained by successive applications of the  $\tilde Q^\dagger$ on the 
vacuum state, to yield $128$ bosons and $128$ fermions
\be \left\{1\ ,\tilde Q^\dagger\ ,(\tilde Q^\dagger)^2\ ,\dots\ ,(\tilde 
Q^\dagger)^7\ ,(\tilde Q^\dagger)^8\right\}\vert~0>\ .\ee
The equality between the number of bosons and fermions is manifest.  
All three irreps have the same quadratic Casimir invariant, since they have the same $I_2/I_0$ ratio. 

Surprisingly, we have found that some higher spin representations of $SO(9)$ also occur in triples with  the same quadratic Casimir invariant, and show remarkable group-theoretical kinships with the supergravity triplet.  The higher spin  triplets appear in four different types, $S-T-T$, $S-S-S$, $T-T-S$, and $T-S-T$, where $S$ describes fermionic (odd $a_4$), and $T$ bosonic  (even $a_4$) degrees of freedom. The largest representation is listed first, and its dimension is equal to the sum of dimensions of the other two irreps of the triplet. Thus only triplet of the $S-T-T$ type display supersymmetry-like properties.
\subsection{S-T-T Triples}
In Dynkinese, these triples are of the form 
\bea [1+p+2r,n,p,1+&2q&+2r]~\oplus ~[2+p+2q+2r,n,p,2r]~\\ \nonumber 
&\oplus& ~[p,n,1+p+2r,2q] 
\eea
labelled by four integers, $n,p,q,r=1,2,\dots$; the sum of  the Dynkin 
invariants 
$I_0$, and $I_2\ ,~I_4\ , ~I_6$, over the bosons match those of the fermion representation. 
All three have the same quadratic Casimir invariant. The simplest of this class is the supergravity multiplet which we have already discussed, and only the lowest of these triples has manifest supersymmetry. The number of fermions and bosons 
of each triplets are equal, and a multiple of $128$, but their construction does not  follow 
that of the supergravity multiplet as polynomials of $\tilde Q^\dagger$ acting 
on some state.  They  appear to be supersymmetric without supersymmetry, the simplest being:
\vskip .5cm
$\bullet$ The supertriple, with $n=1$, $p=q=r=0$,  contains 
\be [2100]_b+[0110]_b+[1101]_f\ ,\ee
with group-theoretic numbers given by 
\hskip 2cm
\begin{center}
\begin{tabular}{|c|c|c|c|}
\hline
$~{\rm irrep}~$&$ [2100]$ & $[0110]$ & $[1101]$  \\
 \hline \hline         
$~I_0~ $&$  910$ & $1650$ & $ 2560$  \\
 \hline    
$~I_2~$& $3640$& $6600$ & $10240$  \\
\hline
$~I_4~$& $19864$& $34920$ & $54784$ \\                                                            
 \hline
$~I_6~$&$130840$& $217320$ &$348160$\\ 
\hline
$~I_8~$&$977944$& $1498344$ &$2466304$\\ 
\hline
 \end{tabular}\end{center}
\vskip 0.3cm
and described by fields of the form
\be h_{(ijk)l}+A_{(ij)(kl)m}+\Psi_{\alpha (ij)k}\ .\ee 
Their index structure indicates the appearance of higher spin fields. 
It is not possible to generate this triple  by repeated use of the light-cone 
supersymmetry algebra 
acting on some  field $\vert~\lambda>$, with dimension equal to $20$
\be \left(1\ ,\tilde Q^\dagger\ ,(\tilde Q^\dagger)^2\ ,\dots\ ,(\tilde 
Q^\dagger)^7\ ,(\tilde Q^\dagger)^8\right)\vert~\lambda>\ .\ee
This would imply that $\vert~\lambda>$ appears twice in the triple, but the triple 
contains no duplicate representations of  $SO(6)\times SO(3)$ that add up to  dimension equal to $20$. These fields  appear in the Kronecker product of the supergravity triplet with the two-form $[0100]$,  
\be
[2000]\otimes [0100]=[2100]\oplus [2000]\oplus \left\{[1010]\oplus 
[0100]\right\}\ ,\ee
\bea
[0010]\otimes [0100]=&[&0110]\oplus [0010]\oplus \\ \nonumber
\{[1002]&\oplus& [1000]\oplus 
[0110]\oplus [1100]\oplus [0002]\}\ ,\eea
\bea
[1001]\otimes [0100]=&[&1101]\oplus [1001]\oplus \\ \nonumber
\{[2001]&\oplus &[1101]\oplus [1001]\oplus 
[0101] \oplus [0011]\oplus [0001]\}\ .\eea
A suitable product can be found that automatically traces out the extra 
representations (in the curly brackets), subtracts all traces, and all totally antisymmetric tensors. The states of this triple could be understood as 
some sort of bound state between the supergravity states and something having the Lorentz 
properties of a $2$- or $7$-form in the light cone little group. Whether this union can be consumated through actual dynamics  remains to be seen. Although this  
feature usually associated with supersymmetry remain, it is not a supermultiplet. 

$\bullet$ The second tower of triples is obtained by multiplying all its representations by $[1010]$, and performing suitable subtractions. This representation appears in the antisymmetric product of two second-rank antisymmetric  tensor fields. It could therefore be generated by applying two two-forms on the supergravity multiplet, resulting in a bound state between supergravity and two branes. The simplest with $p=1$, $n=q=r=0$, contains
\be [3010]_b+[1020]_b+[2011]_f\ ,\ee 
with group theory table
\hskip 2cm
\begin{center}
\begin{tabular}{|c|c|c|c|}
\hline
$~{\rm irrep}~$&$ [3010]$ & $[1020]$ & $[2011]$  \\
 \hline \hline         
$~I_0~ $&$  7700$ & $12012$ & $19712$  \\
 \hline    
$~I_2~$& $46200$& $72072$ & $118272$  \\
\hline
$~I_4~$& $384360$& $585624$ & $969984$ \\                                                            
 \hline
$~I_6~$&$3938760$& $5748792$ &$9687552$\\ 
\hline
$~I_8~$&$46646664$& $64127736$ &$110529792$\\ 
\hline
 \end{tabular}\end{center}
\vskip .5cm
$\bullet$ The third tower is more complicated,  multiplying  the fermion and the second boson by $[0002]$, and the first boson  by $[2000]$. The simplest in this series, with $q=1$ $n=p=r=0$,  and contains 
\be [4000]_b+[0012]_b+[1003]_f\ee 
\hskip 2cm
\begin{center}
\begin{tabular}{|c|c|c|c|}
\hline
$~{\rm irrep}~$&$ [4000]$ & $[0012]$ & $[1003]$  \\
 \hline \hline         
$~I_0~ $&$450$ & $4158$ & $4608$  \\
 \hline    
$~I_2~$& $2200$& $20328$ & $22528$  \\
\hline
$~I_4~$& $15160$& $131784$ & $146944$ \\                                                            
 \hline
$~I_6~$&$130360$& $1016520$ &$1146880$\\ 
\hline
$~I_8~$&$1325944$& $8839560$ &$10103296$\\ 
\hline
 \end{tabular}\end{center}
\vskip 0.5cm
$\bullet$ The fourth infinite tower is also twisted.  The simplest of this series has $r=1$, with content 
\be [4002]_b+[0030]_b+[3003]_f\ ,\ee
and group-theory mugshot
\hskip 2cm
\begin{center}
\begin{tabular}{|c|c|c|c|}
\hline
$~{\rm irrep}~$&$ [4002]$ & $[0030]$ & $[3003]$  \\
 \hline \hline         
$~I_0~ $&$ 32725$ & $23595$ & $56320$  \\
 \hline    
$~I_2~$& $261800$& $188760$ & $450560$  \\
\hline
$~I_4~$& $2938280$& $2055768$ & $4994048$ \\                                                            
 \hline
$~I_6~$&$41127080$& $27239256$ &$68366336$\\ 
\hline
$~I_8~$&$673801256$& $414212568$ &$1084279808$\\ 
\hline
 \end{tabular}\end{center}
\vskip 0.3cm
\subsection{S-S-S Triples}
Here all three representations are spinors
\bea [2+p+2r,n,p,3+&2q&+2r]\oplus 
[4+p+2q+2r,n,p,1+2r]\\ \nonumber
&\oplus&[p,n,2+p+2r,1+2q]\ ,\eea
and the dimension of the first is the sum of the other two. The simplest example is 
\be [2003]\oplus [4001]\oplus [0021]\ .\ee
Its group-theory mugshot is
\hskip 2cm
\begin{center}
\begin{tabular}{|c|c|c|c|}
\hline
$~{\rm irrep}~$&$ [2003]$ & $[4001]$ & $[0021]$  \\
 \hline \hline         
$~I_0~ $&$18480$ & $5280$ & $13200$  \\
 \hline    
$~I_2~$& $117040$& $33440$ & $83600$  \\
\hline
$~I_4~$& $1010992$& $297632$ & $713360$ \\                                                            
 \hline
$~I_6~$&$10640944$& $3303584$ &$7337360$\\ 
\hline
$~I_8~$&$128166448$& $43030688$ &$85922192$\\ 
\hline
 \end{tabular}\end{center}
\vskip 0.3cm
\subsection{T-T-S Triples}
In this class, the dimension of the largest 
boson (listed first) is equal to that of the spinor and the second boson,  
\bea[1+p+2r,n,p,2+&2q&+2r]~\oplus ~[3+p+2q+2r,n,p,2r]\\ \nonumber
&\oplus [&p,n,1+p+2r,1+2q]\ .\eea
The lowest member of this class is
\be
[1002]~\oplus~[3000]~\oplus~[0011]\ ,\ee
with mugshot
\hskip 2cm
\begin{center}
\begin{tabular}{|c|c|c|c|}
\hline
$~{\rm irrep}~$&$ [1002]$ & $[3000]$ & $[0011]$  \\
 \hline \hline         
$~I_0~ $&$924$ & $156$ & $768$  \\
 \hline    
$~I_2~$& $3080$& $520$ & $2560$  \\
\hline
$~I_4~$& $13400$& $2392$ & $11008$ \\                                                            
 \hline
$~I_6~$&$68216$& $13432$ &$54784$\\ 
\hline
$~I_8~$&$382328$& $87544$ &$299776$\\ 
\hline
 \end{tabular}\end{center}
\vskip 0.3cm
\subsection{T-S-T Triples}
The last class contains the representations
\bea[2+p+2r,n,p,2+&2q&+2r]~\oplus ~[3+p+2q+2r,n,p,1+2r]\\ \nonumber
&\oplus [&p,n,2+p+2r,2q]\ .\eea
Its lowest-lying member is
\be
[2002]~\oplus~[3001]~\oplus~[0020]\ ,
\ee
with mugshot
\hskip 2cm
\begin{center}
\begin{tabular}{|c|c|c|c|}
\hline
$~{\rm irrep}~$&$ [2002]$ & $[3001]$ & $[0020]$  \\
 \hline \hline         
$~I_0~ $&$3900$ & $1920$ & $1980$  \\
 \hline    
$~I_2~$& $18200$& $8960$ & $9240$  \\
\hline
$~I_4~$& $114920$& $57728$ & $57192$ \\                                                            
 \hline
$~I_6~$&$875720$& $455936$ &$419784$\\ 
\hline
$~I_8~$&$7549064$& $4148096$ &$3453384$\\ 
\hline
 \end{tabular}\end{center}
\vskip 0.3cm
There are several  triples which only match dimensions and 
quadratic Casimir invariants; we found one made entirely of spinors
\be [1033]\oplus [7001]\oplus [0305]\ ;\qquad 
[7122]\oplus [6008]\oplus [4018]\ ,\ee
with the dimension of the first equal to the sum of the other two, and 
all with the same quadratic Casimir, buth their $I_{4,6}$ do not match. 
\subsection{Basic Operations}
It is possible to understand these different triples in terms of four basic 
operations, which starting from the supergravity multiplet, generate all 
triples:
\begin{itemize}
\item $\Delta_1$: Increase the Dynkin labels all three irreps within a 
triple by $[0100]$.
\item $\Delta_2$: Increase the Dynkin labels of all three irreps within a 
triple by $[1010]$.
\item $\Delta_3$: Increase the Dynkin labels of the first and third irreps by $[0001]$,  
the second by $[1000]$. 
\item $\Delta_4$: Increase the Dynkin labels of the first and second irreps by $[1001]$, 
the third by $[0010]$ 
\end{itemize}
 The $\Delta_{1,2}$ operations may be simplest to understand as they can 
be generated by applying  representations that appear either as the light-cone $2$-form $[0100]$, or in its  twice-antisymmetrized  product, since 
\be
\left([0100]\otimes [0100]\right)_A=[0100]\oplus [1010]\ .\ee
A light-cone $2$-form may indicate a brane state, and these triples could 
then be understood as bound states of the supergravity fields with these 
branes. The third and fourth operations are   more complicated as they treat the 
different members differently. However, starting from the supergravity 
multiplet, they generate all other triples, as shown in the diagram 
below, where the upward arrow denotes $\Delta_4$, and the downward arrow 
denotes $\Delta_3$:

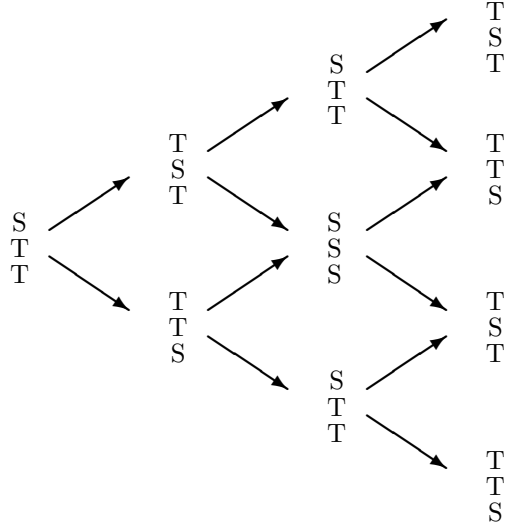
\begin{picture}(200,200)(-100,20)

\put(10,100){\shortstack{S\\T\\T}}
\put(70,70){\shortstack{T\\T\\S}}
\put(70,130){\shortstack{T\\S\\T}}

\put(130,100){\shortstack{S\\S\\S}}
\put(130,40){\shortstack{S\\T\\T}}
\put(130,160){\shortstack{S\\T\\T}}

\put(190,10){\shortstack{T\\T\\S}}
\put(190,70){\shortstack{T\\S\\T}}
\put(190,130){\shortstack{T\\T\\S}}
\put(190,180){\shortstack{T\\S\\T}}

\thicklines
\put(25,120){\vector(3,2){30}}
\put(25,110){\vector(3,-2){30}}

\put(85,150){\vector(3,2){30}}
\put(85,80){\vector(3,-2){30}}
\put(85,90){\vector(3,2){30}}
\put(85,140){\vector(3,-2){30}}

\put(145,180){\vector(3,2){30}}
\put(145,170){\vector(3,-2){30}}
\put(145,120){\vector(3,2){30}}
\put(145,110){\vector(3,-2){30}}
\put(145,60){\vector(3,2){30}}
\put(145,50){\vector(3,-2){30}}
\end{picture}
\vskip 1cm
It is clear that the supergravity multiplet sits at the beginning of a 
very intricate and beautiful complex of irreps of $SO(9)$. Limited by the two 
dimensions of the paper, we have not shown the effect of the $\Delta_{1,2}$ operations 
which act uniformly on any of the triples in the picture. The whole pattern is summarized by the general form of the triples
\vskip .25cm
\bea [1+a_2+a_3,a_1,a_2,&1&+a_3+a_4]~\oplus ~[2+a_2+a_3+a_4,a_1,a_2,a_3]\\ \nonumber
~&\oplus& ~[a_2,a_1,1+a_2+a_3,a_4] \ .\eea
\vskip .25cm
\noindent where $a_i$ are non-negative integers.
\section{Mathematical Origin of the Triples}
So far we have only offered numerical evidence for the remarkable structure of the $SO(9)$ representations. A recent paper~\cite{PNAS} by  B. Gross, B. Kostant, S. Sternberg and one of us (PR), unveils its mathematical origin. The following is a watered-down version of its contents. It points to a construction of a more general character, but does not (yet) seem to shed light on its physical interpretation. 

The triples stem from the triality of $SO(8)$, which is explicitly realized in $F_4$, and 
the three equivalent ways to embed $SO(9)$ into $F_4$. That very triality is already familiar to particle physicists: the three equivalent ways to embed $SU(2)\times U(1)$ in $SU(3)$, called I-spin, U-spin, and V-spin~\cite{Meshkov}. 

The general idea behind the mathematical construction goes as follows. Let F and B be two Lie algebras of equal rank such that $F\supset B$. The Weyl group of F, ${\cal W}(F)$,  is  bigger than that of B, ${\cal W}(B)$, with $r$-times as many operations. The fundamental Weyl chamber of F, is the  sliver of weight space that contains all weights with positive or zero Dynkin labels; it is $r$ times smaller than the fundamental Weyl chamber of the subgroup B. 

Let $\lambda$ be the highest weight of an irrep of F; it lies either inside or at the boundary of the Weyl chamber. We can choose  $r$ 
operations of  ${\cal W}(F)$ not in ${\cal W}(B)$
 which map  the fundamental Weyl chamber of F into that of B. 
When applied to this highest weight, they produce $r$ copies inside
the chamber of B (unless the weight is at the chamber boundary).
  In order to make sure it is inside the chamber, we add to it the weight $\rho=[1,1,1,\dots,1]$, the Weyl vector (or half sum of positive roots). Then we are sure the Weyl group will act on this weight non-trivially. We now construct the $r$ weights
\be
\lambda_i\equiv w_i(\lambda+\rho_F)-\rho_B\ ,\ \ i=1,2\dots r\ ,\ee
where $w_i$ are the operations ${\cal W}(F)$ not in ${\cal W}(B)$. They all lie on the fundamental Weyl chamber of B, and on its boundary, and therefore describe an $r$-plet of irreps of B. 

Apply this reasoning to the case $F_4\supset SO(9)$. The Weyl group of $F_4$ has dimension $1152$, that of $SO(9)$, $384$ so that $r=3$. Starting from any representation of $F_4$ this construction generates a triplet of representations of $SO(9)$. There remains to identify those three elements of the Weyl group, the reason for the relations among their invariants and the emergence of supersymmetry in the construction.

To understand the origin of the triality in $F_4$, the octonion language is convenient since the adjoint of $F_4$ is generated by  antihermitian traceless $3\times 3$ matrices over the octonions, supplemented by their automorphism group, $G_2$. Triality is then related to 
the  three inequivalent ways of picking out one of the matrix's off-diagonal elements, and this construction generalizes to the Lie
algebras of the {\it magic square}~\cite{Intro}.

Octonions, together with  real numbers, R, complex numbers, C, quaternions, Q, are 
the four Hurwitz (division) algebras. $3\times 3$ matrices with elements belonging to these algebras generate interesting mathematical structures. 
\begin{itemize}
\item For real numbers, these matrices generate the Lie algebra $SO(3)$. Its 
maximal subgroup is $SO(2)$.
\item For complex numbers, they generate the Lie algebra $A_2\sim SU(3)$, 
and singling out one of the three off-diagonal elements picks out the 
subgroup $SU(2)\times U(1)\sim SO(3)\times SO(2)$.
\item For quaternions, together with their automorphism group $A_1\sim 
SU(2)$, they generate $C_3\sim Sp(6)$. Two off-diagonal elements are 
treated equally by the subgroup $Sp(4)\times Sp(2)\sim SO(5)\times SO(3)$.
\item With octonions, and their automorphism group $G_2$, they generate 
the exceptional group $F_4$. Its subgroup $B_4\sim SO(9)$ naturally picks 
out one of the three off-diagonal elements.
\end{itemize}
Under $F_4\supset SO(9)$, its adjoint breaks up as
${\bf 52}={\bf 36}+{\bf 16}$, where ${\bf 36}$ is the adjoint of $SO(9)$ and $\bf 16$ its spinor representation. Another way to look at this embedding is to say that it generates a $16$-dimensional coset space acted on by the orthogonal group $SO(16)$. It yields 
the anomalous embedding $SO(16)\supset SO(9)$ according to which the spinor of $SO(9)$ fits in the vector of $SO(16)$. 
\section{The Magic Square}
Starting from the four Hurwitz algebras, it is possible to construct the so-called composition algebras, which include all the exceptional groups, except $G_2$, the automorphism group of the octonions. That construction~\cite{Intro} relies on the triality of both $SO(8)$ and on 
the structure of $3\times 3$ matrices. Start from $3\times 3$ antihermitian traceless matrices with elements over the product of any two of the four Hurwitz algebras, with  three off-diagonal elements, and two diagonal elements which are pure imaginary. They are acted on by 
the automorphism groups of the algebras, and each of their parameters
generate a  Lie algebra transformation, to produce one of the ten Lie
algebras in  the {\it magic square}:
\hskip 3cm
\begin{center}
\begin{tabular}{|c|c|c|c|c|}
\hline
$$&$R$ & $C$ & $Q$&$O$  \\
 \hline \hline         
$~R~ $&$~SO(3)~$ & $~SU(3)~$ & $~Sp(6)~$&$~F_4~$  \\
 \hline    
$~C~$& $~SU(3)~$& $~SU(3)\times SU(3)~$ & $~SU(6)~$ &$~E_6~$ \\
\hline
$~Q~$& $~Sp(6)~$& $~SU(6)~$ & $~SO(12)~$&$~E_7~$ \\                                                            
 \hline
$~O~$&$~F_4~$& $~E_6~$ &$~E_7~$&$~E_8~$\\ 
\hline
\end{tabular}\end{center}
\vskip 0.3cm
In particular, the exceptional group $F_4$ is generated by $3\times 3$ traceless antihermitian matrices over the octonions, together with  $G_2$,  the automorphism group of the octonion 
multiplication table. This produces the $3\times 8+2\times 7+14=52$ parameters of the algebra. 

Each of the algebras appearing in the magic square have subalgebras of equal rank, and we can apply our mathematical construction to each. The results are summarized in table below, first for the exceptional groups,
\hskip 2cm
\begin{center}
\begin{tabular}{|c|c|c|c|}
\hline
${\rm Group}$ & ${\rm Subgroup}$ & ${\rm Coset~Dimension}$ &$r$ \\
 \hline \hline         
$~E_8~ $&$SO(16) $ & $128$ & $135$  \\
 \hline    
$~E_7~$& $SO(12)\times SO(3)$& $64$ & $63$  \\
\hline
$~E_6~$& $SO(10)\times SO(2)$& $32$ & $27$ \\                                                            
 \hline
$~F_4~$&$SO(9)$& $16$ &$3$\\  
\hline
 \end{tabular}\end{center}
\vskip 0.3cm
and then for the non-exceptional groups in the square
\hskip 2cm
\begin{center}
\begin{tabular}{|c|c|c|c|}
\hline
${\rm Group}$ & ${\rm Subgroup}$ & ${\rm Coset~Dimension}$ &$r$ \\
 \hline \hline         
$~SO(12)~ $&$SO(8)\times SO(4)$ & $32$ & $30$  \\
 \hline    
$~SU(6)~$& $SO(6)\times SO(3)\times SO(2)$& $16$ & $15$  \\
\hline
$~Sp(6)~$& $Sp(4)\times Sp(2)$& $8$ & $3$ \\                                                            
 \hline
$~SU(3)\times SU(3)~$&$SO(3)\times SO(3)\times SO(2)\times SO(2)$& $16$ &$3$\\  
\hline
$~SU(3)~$&$SO(3)\times SO(2)$& $4$ &$3$\\  
\hline
 \end{tabular}\end{center}
\vskip 0.3cm
The connection with supersymmetry occurs through the generation of representations of the subgroup in terms of polynomials in Clifford charges. 
\subsection{$SU(3)$ Triples}
By studying the simplest non-trivial example, of the embedding of $SU(2)\times U(1)$ into $SU(3)$, the emergence of supersymmetry in our general construction should become clear.   We begin with some well known facts about $SU(3)$, the algebra 
generated by $3\times 3$ antihermitian traceless matrices over the 
complex numbers. Its maximal subgroup is 
\be SU(3)\supset SU(2)\times U(1)\sim SO(3)\times SO(2)\ .\ee 
There are three equivalent ways to imbed this subalgebra in $SU(3)$, 
corresponding, in particle physics language, to $I$-spin, $U$-spin and 
$V$-spin. These three embeddings can be understood in terms of 
the Weyl group. The Weyl group of $SU(3)$ is $S_3$, the permutation 
group on three objects. It is three times as big as the Weyl group of its 
maximal subgroup $SO(3)\times SO(2)$.

All the  states spanned by this algebra live in a 
$2$-dimensional lattice. In the Dynkin basis, the weights are labelled by 
two integers $[a_1,a_2]$. The action of the Weyl group is simplest acting on an 
orthonormal basis $\{e_i\}$ in which the roots have components which are 
half-integers between $-2$ and $2$.  The simple roots of $SU(3)$ are given in terms of the 
three vectors $e_a$ which span a three-dimensional the Euclidean space
\be
\alpha_1=e_1-e_2\ ;\qquad\alpha_2=e_2-e_3\ .
\ee
A weight is labelled in the orthonormal basis by three numbers, $(b_1,b_2,b_3)$, such that 
$b_1+b_2+b_3=0$. It can also be expressed in the Dynkin basis as
\be w=a_1\omega_1+a_2\omega_2\ ,\ee
where the fundamental vectors $\omega_{1,2}$ are determined through
\be
\label{bases} 2{(\alpha_i,\omega_j)\over (\alpha_i,\alpha_i)}=\delta_{ij}\ .\ee
It follows that the same weight has the orthonormal basis components
\be
b_1={1\over 3}(2a_1+a_2)\ ,\ b_2={1\over 3}(a_2-a_1)\ , \ b_3=-{1\over 3}(a_1+2a_2) \ .\ee
The action of the Weyl group on the orthonormal components is just the permutation group on 
the three  $b_i$'s. Furthermore,  the fundamental Weyl chamber is described by  the 
inequalities $b_1>b_2>b_3$. We will use the Weyl vector, defined as half the sum of the positive roots: 
\be \rho^{}_{SU(3)}=(1,0 ,-1 )\sim [1,1]\ ,\ee
where the square brackets in indicate the Dynkin basis and the curved brackets the orthonormal basis.

Start from an irrep of $SU(3)$, labelled by its maximum 
Dynkin weight $[a_1,a_2]$, or orthonormal components $(b_1,b_2,b_3)$, and add the Weyl vector, so that the resulting weight is inside the 
chamber.  Since the Weyl group of $SU(3)$ is three times as large as that of $SU(2)$, the fundamental Weyl chamber of $SU(2)$, is three times as large as that of $SU(3)$. Identify the three Weyl transformations (permutations)  that produce a weight in the fundamental Weyl chamber of $SU(2)$, defined by $b'_1\ge b'_2$. These yield three weights 
\be
(b_1+1,b_2,b_3-1)\qquad(b_1+1,b_3-1,b_2)\qquad (b_2,b_3-1,b_1+1)\ .\ee
Next we subtract the $SU(2)$ Weyl vector, $\rho_{SU(2)}=(1/2,-1/2)\sim [1]$, and revert to the Dynkin basis. This 
construction yields the $SU(3)$-generated triples, with the $U(1)\sim SO(2)$ 
charge indicated as a subscript
\vskip .25cm
\be
SU(3):~~~~~~[a_1]_{{1\over 2}(a_1+1)}~\oplus~[a_1+a_2+1]_{{1\over 
2}(a_1-a_2)}
~\oplus~[a_2]_{-{1\over 2}(a_2+1)}\ee
\vskip .25cm
The three Weyl operations are best identified by their action on the fundamental irrep of $SU(3)$, $u,d,s$: 
\bea
w_1&:&~(u,d,s)\rightarrow (u,d,s)\ ,\\
 w_2&:&~(u,d,s)\rightarrow (u,s,d)\ ,\\
 w_3&:&~ (u,d,s)\rightarrow (s,u,d)\ .\eea
Starting from the singlet of $SU(3)$, with 
 $a_1=a_2=0$, we get the simplest triple
\be
[0]_{1\over 2}\oplus [1]_{0}\oplus[0]_{-{1\over 2}}\ .\ee
This triplet of representations forms a supersymmetric multiplet, generated by a charge that is a doublet under the $SU(2)$ and has $U(1)$ value of $-1/2$. To understand its origin, note that the group acting on the four-dimensional coset $SU(3)/SU(2)\times U(1)$ is $SO(4)\equiv SO(3)\times SO(3)$,  defining an embedding of $SO(4)\supset SU(2)\times U(1)$. The states of this lowest triple correspond to the decomposition of the  $SO(4)$ spinor.  

An obvious interpretation is to view $SO(3)\times SO(2)$ as the compact subgroup as the light cone little group of either $SO(4,1)\times SO(2)$, or $SO(3)\times SO(3,1)$, which lead to 
theories in $d=5$ and $d=4$ dimensions, respectively. 

Two fundamental operations generate the higher triples, in one to one 
correspondance with the rank of the mother algebra $SU(3)$. 
The operation that increases $a_1$ by two, produces on $[0]_{1\over 2}$ the infinite chain
\be
[0]_{1\over 2}~\oplus~[2]_{3\over 2}~\oplus~[4]_{5\over 2}~\oplus~\cdots\ .\ee
It is amusing that these states fit in the spin zero singleton 
(Rac) representation of $SO(3,2)$. The same operation on $[1]_0$ yields the Di representation. The $SO(2)$ charge is the energy and 
the $SO(3)$ representation determines the spin.
\subsection{$Sp(6)$ Triples}
We now apply the same construction to the next entry in the magic square. The Lie algebra $Sp(6)\equiv C_3 $ is generated by the $3\times 3$ traceless antihermitian matrices over quaternions, augmented by $SU(2)$, the automorphism group of the quaternions. By picking out one of its off-diagonal element, we obtain the embedding
\be
Sp(6)\supset Sp(4)\times Sp(2)\sim SO(5)\times SO(3)\ ,\ee
with an eight-dimensional coset space acted on by $SO(8)$. $Sp(6)$ has three simple roots, given by
\be
\alpha_1=e_1-e_2\ , \ 
\alpha_2=e_2-e_3\ ,\ 
\alpha_3=2e_3\ .\ee
 The same weight, written in  Dynkin and orthonormal bases, has components 
$[a_1,a_2,a_3]$ and $(b_1,b_2,b_3)$, respectively with the relations
\be
b_1=a_1+a_2+a_3\ , \ b_2=a_2+a_3\ , \ b_3=a_3\ ,\ee
derived from Equation (\ref{bases}), where now $i,j=1,2,3$, 
after carefully accounting for the long root. Its Weyl vector is given by
\be
\rho_{C_3}=(3,2,1)\sim[1,1,1]\ee
The Weyl group of $C_3$ contains $S_3$, which permutes the three $b_i$'s. As in the previous case, three of its operations relate the three equivalent ways to embed $SO(5)\times SO(3)$. 

Add to the  highest weight of an irrep of $C_3$, $[a_1,a_2,a_3]$ the Weyl vector, 
to put it inside the fundamental chamber of $Sp(6)$. There are  three elements of 
the Weyl group which map this weight into the fundamental chamber of the subgroup $SO(5)\times SO(3)$, defined by $b_1\ge b_2\ge 0$, and $b_3\ge 0$: the identity element, the parity $P_{23}$,  which interchanges $b_2$ and $b_3$,  and the cyclic element $C_{231}$ which effects $(b_1,b_2,b_3)\rightarrow (b_2,b_3,b_1)$. After  subtracting the Weyl vector of the subgroup
$ \rho_{}=(2,1;1)$, where the last entry is for the $SO(3)$ subgroup, we obtain the three weights
\be(b_1+1,b_2+1;b_3)\ ;\qquad (b_1+1,b_3;b_2+1)\ ;\qquad (b_2,b_3;b_1+2)\ .\ee
To rewrite these in the Dynkin basis, we use the expression for the simple roots of $SO(5)$ in the orthonormal basis
\be\alpha_1=e_1-e_2\ ,\qquad \alpha_2=e_2\ ,\ee
 which yield the $Sp(6)$-generated triple
\vskip .25cm
\bea
[a_1,a_2+a_3+&1&;a_3]~\oplus~[a_1+a_2+1,a_3;a_2+a_3+1]\\ \nonumber 
~&\oplus&~[a_2,a_3;a_1+a_2+a_3+2]\eea
\vskip .25cm
The scalar irrep of $Sp(6)$ yields the simplest triplet
\be
[0,1;0]~\oplus~[1,0;1]
~\oplus~[0,0;2]\ ,\ee
or in terms of the dimensions of the representations of 
$SO(5)\times SO(3)$, 
\be
({\bf 5},{\bf 1})\oplus~({\bf 4},{\bf 2})\oplus~({\bf 1},{\bf 3})\ ,\ee
the representation content of super-Yang-Mills in ten dimensions. The orthogonal group $SO(8)$ acts on the eight-dimensional coset space, defining an embedding of the spinor of $SO(5)\times SO(3)$ into the vector irrep of $SO(8)$: ${\bf 8}_V=({\bf 4},{\bf 2})$. The triple is just the decomposition of the spinors of $SO(8)$: ${\bf 8}_S=({\bf 4},{\bf 2})$, ${\bf 8}_S'=({\bf 5},{\bf 1})+({\bf 1},{\bf 3})$. The subgroup can be interpreted as the light-cone little group of either $SO(6,1)\times SO(3)$, or $SO(5)\times SO(4,1)$, which imply 
Lorentz-invariant theories in either $d=7$ or $d=5$ space-time dimensions.
\subsection{$F_4$ Triples}
The $F_4$ simple roots are given by
\bea
\alpha_1&=&e_2-e_3\ ,\qquad
\alpha_2=e_3-e_4\ ,\\
\alpha_3&=&e_4\ ,\qquad ~~~~~~~
\alpha_4={1\over 2}(e_1-e_2-e_3-e_4)
\ .\eea
Through the use of Equation (\ref{bases}), we find the relation between the 
components of any weight in the Dynkin basis $[a_1,a_2,a_3,a_4]$  and the orthonormal 
 basis $(b_1,b_2,b_3,b_4)$ 
\bea
b_1&=&a_1+2a_2+{3\over 2}a_3+a_4\ ,\ 
b_2=a_1+a_2+{1\over 2}a_3\ ,\\ 
b_3&=&a_2+{1\over 2}a_3\ ,\ b_4={1\over 2}a_3\ .\eea
The Weyl vector is given in the orthonormal and Dynkin bases by
\be
\rho^{}_{F_4}=~{1\over 2}(11,5,3,1)~\sim~[1,1,1,1]\ .\ee

To generate the triples, we start with an irrep of $F_4$, $[a_1,a_2,a_3,a_4]$, and add to it 
the Weyl vector to put it inside the fundamental chamber. Express it in the orthonormal basis. 
The three elements of the $F_4$ Weyl group which map weights inside the fundamental chamber of $SO(9)$, are the identity, a parity and an anticyclic permutation. These are most easily expressed by their action on the fundamental of $F_4$, which is a $3\times 3$ hermitian traceless octonionic matrix. The parity interchanges two off-diagonal octonion elements, and the cyclic permutes the three off-diagonal octonion elements. The first is the original 
weight, the second and third are obtained in terms of permutations
on the simple roots of $D_4$, for which we have
\bea
\alpha_1&=&e_1-e_2\ ,\qquad 
\alpha_2=e_2-e_3\ ,\\
\alpha_3&=&e_3-e_4\ ,\qquad
\alpha_4=e_3+e_4
\ .\eea
In this numbering, $\alpha_2$ is at the center of the Dynkin diagram, and 
does not move under the permutations. The two permutations that produce 
the required weights are 
\bea
S_{14}&:&~~~\alpha_1\mapsto \alpha_4\ ,~\alpha_4\mapsto \alpha_1\ 
,~\alpha_3\mapsto \alpha_3\ ,\\
C_{143}&:&~~~\alpha_1\mapsto \alpha_4\mapsto \alpha_3\mapsto\alpha_1\ .\eea
Weights in the chamber of $B_4$ are determined by the inequalities $b_1\ge b_2\ge b_3\ge b_4\ge 0$. Next, to get in the closure of the Weyl chamber, we subtract the Weyl vector of $SO(9)$,
\be
\rho_{B_4}={1\over 2}(7,5,3,1)\sim[1,1,1,1]\ ,\ee
and convert in the Dynkin basis of $SO(9)$, using  the $B_4$ relations
\bea
\alpha_1&=&e_1-e_2\ ,\qquad
\alpha_2=e_2-e_3\ ,\\
\alpha_3&=&e_3-e_4\ ,\qquad
\alpha_4=e_4\ ,\eea
from which 
\bea
a_1&=&b_1-b_2\ ,\qquad
a_2=b_2-b_3\ ,\\
a_3&=&b_3-b_4\ ,\qquad
a_4=2b_4\ .\eea
This leads back to the $F_4$ generated triples in the Dynkin basis
\vskip .25cm
\bea
[1+a_2+a_3,a_1,a_2,1+&a_3&+a_4]~\oplus~[2+a_2+a_3+a_4,a_1,a_2,a_3]\\ \nonumber
~\oplus~&[&a_2,a_1,1+a_2+a_3,a_4]\ .\eea
\vskip .25cm
This formula corresponds to that previously derived empirically.
For the simplest case, the triple is the supergravity multiplet
\be
[1,0,0,1]~\oplus~[2,0,0,0]
~\oplus~[0,0;1,0]\ .\ee
The group $SO(16)$ acts on the sixteen-dimensional coset, providing the embedding of the spinor of $SO(9)$ into the vector of $SO(16)$. Then the lowest order triple is the decomposition of the two spinors of $SO(16)$ into $SO(9)$, which explains the supersymmetry.  
\subsection{$E_{6,7,8}$-Multiples}
As indicated by the decompositions of the $E$-like exceptional groups in the magic square,
we can generate higher order Clifford algebras. 
\begin{itemize}
\item The embedding $E_6\supset SO(10)\times SO(2)$ produces a $32$-dimensional coset space. The lowest multiplet contains $27$ terms which are decomposition of the two spinor irreps of $SO(32)$ in terms of $SO(10)\times SO(2)$. These are generated by a Clifford algebra with $2^{16}$ elements, half fermions, half bosons. The $SO(32)$ Clifford is also generated by the magic square embedding $SO(12)\supset SO(8)\times SO(4)$. 
\item With $E_7\supset SO(12)\times SO(3)$, a $64$-dimensional coset space is generated. The two spinor irreps of $SO(64)$ break into 63 irreps of the subgroup, with $2^{32}$ states generated by a Clifford algebra, again half of them fermions.
\item Finally, $E_8\supset SO(16)$ produces  $2^{64}$ states describing the two spinors of $SO(128)$ in terms of $135$ irreps of $SO(16)$!  
\end{itemize}
In all these cases, the lowest multiplet that is generated contains states of spin higher than two, which means that any theory based on these structures probably have 
no local limit, except perhaps by compactification.

Finally we note that the magic square embedding $SU(6)\supset SO(6)\times SO(3)\times SO(2)$ yields in the lowest multiplet the states generated by the $SO(16)$-Clifford, corresponding to the dimensional reduction of the sugra multiplet, and it could be viewed as coming from $d=8-,5-,4-$dimensional theories or even on $AdS_4\times S_6$. For the reader interested in proofs concerning the properties of these multiplets, we refer the reader to ~\cite{PNAS}, where a novel formula between representations of Lie algebras of the same rank is derived. 

In conclusion, we have described a very rich mathematical structure, which contains as special cases, the supermultiplets of eleven-dimensional supergravity and ten-dimensional 
super-Yang-Mills. While there is at present no clear interpretation of these structures, it is tempting to believe that they may shed some light on the eventual structure of $M$-theory. 
\section{Acknowledgements} 
One of us (PR) acknowledges the hospitality of the Aspen Center for Physics and illuminating conversations with Profs L. Brink, R. Moody and S. Sternberg. 
This work is supported in part by the
United States Department of Energy under grant DE-FG02-97ER41029.

\end{document}